\documentclass{elsart}

\usepackage{graphicx}
\usepackage{amssymb}

\begin{document}

\begin{frontmatter}

\title{Magnetic properties of cuprate perovskites\\ in the normal state}

\author{A.~Sherman}

\ead{alexei@fi.tartu.ee}

\address{Institute of Physics, University of Tartu, Riia 142, 51014
Tartu, Estonia}

\begin{abstract}
Normal-state magnetic properties of cuprate high-$T_c$ superconductors
are interpreted based on the self-consistent solution of the $t$-$J$
model of Cu-O planes. The solution method retains the rotation symmetry
of spin components in the paramagnetic state and has no preset magnetic
ordering. The obtained solution is homogeneous. The calculated
temperature and concentration dependencies of the magnetic
susceptibility are close to those observed in experiment. These results
offer explanations for the observed scaling of the static uniform
susceptibility and for the changes in the spin correlation length,
spin-lattice and spin-echo decay rates in terms of the temperature and
doping variations in the spin excitation spectrum.
\end{abstract}


\end{frontmatter}

Magnetic properties of cuprate perovskites have been extensively
studied during the last years, both because of their unusual behavior
and in the hope that they might provide insight into the physical
origin of high-$T_c$ superconductivity
\cite{Johnston,Imai,Takigawa,Chubukov,Barzykin,Jaklic,Fong}.
Considerable progress has been made in this field with the use of
phenomenological approaches, exact diagonalization of small clusters
and -- for heavily overdoped materials -- in a RPA treatment. However,
the basic issues of the magnetic behavior of underdoped cuprates, where
strong electron correlations reveal themselves in full measure, have
not yet been completely resolved. In this paper magnetic properties of
underdoped cuprates in the normal state are investigated using the
$t$-$J$ model of Cu-O planes and the method of Ref.~\cite{Sherman02}
for calculating Green's functions. This method has merits of retaining
the rotation symmetry of spin components in the paramagnetic state and
of the absence of any predefined magnetic ordering. Test calculations
with this method for small clusters and for the undoped case
demonstrated good agreement with the exact diagonalization and Monte
Carlo results. For the considered 20$\times$20 lattice and parameters
of cuprates the calculated magnetic properties of the model appear to
be close to those observed in these crystals. The results give an
insight into mechanisms responsible for the unusual magnetic behavior
of cuprate perovskites.

The Hamiltonian of the 2D $t$-$J$ model reads \cite{Dagotto}
\begin{equation}
H=\sum_{\bf nm\sigma}t_{\bf nm}a^\dagger_{\bf n\sigma}a_{\bf
m\sigma}+\frac{1}{2}\sum_{\bf nm}J_{\bf nm}\left(s^z_{\bf n}s^z_{\bf
m}+s^{+1}_{\bf n}s^{-1}_{\bf m}\right), \label{hamiltonian}
\end{equation}
where $a_{\bf n\sigma}=|{\bf n}\sigma\rangle\langle{\bf n}0|$ is the
hole annihilation operator, {\bf n} and {\bf m} label sites of the
square lattice, $\sigma=\pm 1$ is the spin projection, $|{\bf
n}\sigma\rangle$ and $|{\bf n}0\rangle$ are site states corresponding
to the absence and presence of a hole on the site. For nearest neighbor
interactions $t_{\bf nm}=-t\sum_{\bf a}\delta_{\bf n,m+a}$ and $J_{\bf
nm}=J\sum_{\bf a}\delta_{\bf n,m+a}$ where $t$ and $J$ are the hopping
and exchange constants and the four vectors {\bf a} connect nearest
neighbor sites. The spin-$\frac{1}{2}$ operators can be written as
$s^z_{\bf n}=\frac{1}{2}\sum_\sigma\sigma|{\bf
n}\sigma\rangle\langle{\bf n}\sigma|$ and $s^\sigma_{\bf n}=|{\bf
n}\sigma\rangle\langle{\bf n},-\sigma|$.

The method suggested in Ref.~\cite{Sherman02} is based on Mori's
projection operator technique \cite{Mori} which allows one to represent
Green's functions in the form of continued fractions and gives a way
for calculating their elements. The residual term of the fraction can
be approximated by the decoupling which reduces this many-particle
Green's function to a product of simpler functions. In this way we
obtained the following self-energy equations for the hole $G({\bf k}t)=
-i\theta(t)\langle\{a_{\bf k\sigma}(t),a^\dagger_{\bf k\sigma}\}
\rangle$ and spin $D({\bf k}t)=-i\theta(t)\langle[s^z_{\bf k}(t),
s^z_{\bf -k}]\rangle$ Green's functions:
\begin{eqnarray}
D({\bf k}\omega)&=&\frac{[4J\alpha(\Delta+1+\gamma_{\bf
k})]^{-1}\Pi({\bf k}\omega)+4JC_1(\gamma_{\bf k}-1)}{\omega^2-\Pi({\bf
k}\omega)- \omega^2_{\bf k}}, \nonumber\\
&&\label{se} \\
G({\bf k}\omega)&=&\phi[\omega-\varepsilon_{\bf k}+\mu-\Sigma({\bf
k}\omega)]^{-1}, \nonumber
\end{eqnarray}
where $\gamma_{\bf k}=\frac{1}{4}\sum_{\bf a}\exp(i{\bf ka})$, $\mu$ is
the chemical potential, $\phi=\frac{1}{2}(1+x)$, $x$ is the hole
concentration and
\begin{eqnarray}
\omega^2_{\bf k}&=&16J^2\alpha|C_1|(1-\gamma_{\bf k})
(\Delta+1+\gamma_{\bf k}),\nonumber\\
&&\label{seed}\\
\varepsilon_{\bf k}&=&(4\phi
t+6C_1\phi^{-1}t-3F_1\phi^{-1}J)\gamma_{\bf k}.\nonumber
\end{eqnarray}

The parameter of vertex correction $\alpha$, which improves the
decoupling in the residual term, is set equal to its value in the
undoped case, $\alpha=1.802$. The parameter $\Delta$ which describes a
gap in the spectrum of spin excitations at $(\pi,\pi)$ [see
Eq.~(\ref{seed})] is determined by the constraint of zero site
magnetization $\langle s^z_{\bf l}\rangle=0$ in the paramagnetic state.
The constraint can be written in the form
\begin{equation}\label{zsm}
\frac{1}{2}(1-x)=\frac{2}{N}\sum_{\bf k}\int_0^\infty d\omega
\coth\!\left(\frac{\omega}{2T}\right)B({\bf k}\omega),
\end{equation}
where $B({\bf k}\omega)= -\pi^{-1}{\rm Im}\,D({\bf k}\omega)$ is the
spin spectral function, $N$ is the number of sites and $T$ is the
temperature. Notice that in the considered 2D system the long-range
antiferromagnetic ordering is destroyed at any nonzero $T$
\cite{Mermin} and, as can be seen from the above formulas, at $T=0$ and
$x \gtrsim 0.02$. The value of $x$ and the nearest neighbor
correlations $C_1=\langle s_{\bf l}^{+1}s_{\bf l+a}^{-1}\rangle$ and
$F_1=\langle a_{\bf l}^\dagger a_{\bf l+a}\rangle$ are defined from
$G({\bf k}\omega)$ and $D({\bf k}\omega)$ in the usual way.

The self-energies in Eq.~(\ref{se}) read
\begin{eqnarray}
{\rm Im}\,\Pi({\bf k}\omega)&=&\frac{16\pi t^2J}{N}(\Delta+1+
 \gamma_{\bf k})\sum_{\bf k'}(\gamma_{\bf k}-\gamma_{\bf
 k+k'})^2 \nonumber\\
&&\times \int^\infty_{-\infty}d\omega'[n_F(\omega+\omega')-
 n_F(\omega')]A({\bf k+k'},\omega+\omega')A({\bf k'}\omega'),
 \nonumber\\
{\rm Im}\,\Sigma({\bf k}\omega)&=&\frac{16\pi t^2}{N\phi}\sum_{\bf k'}
 \int_{-\infty}^\infty d\omega'[n_B(-\omega')+n_F(\omega-\omega')]
 \label{po}\\
&&\times \left[\gamma_{\bf k-k'}+\gamma_{\bf
 k}+{\rm sgn}(\omega')(\gamma_{\bf
 k-k'}-\gamma_{\bf k})\sqrt{\frac{1+\gamma_{\bf k'}}{1-\gamma_{\bf
 k'}}}\right]^2 \nonumber\\
&&\times A({\bf k-k'},\omega-\omega')B({\bf k'}\omega'), \nonumber
\end{eqnarray}
where $n_F(\omega)=[\exp(\omega/T)+1]^{-1}$, $n_B(\omega)=\left[
\exp(\omega/T)-1\right]^{-1}$ and $A({\bf k}\omega)=-\pi^{-1}{\rm
Im}\,G({\bf k}\omega)$ is the hole spectral functions. The source of
damping of spin excitations described by Eq.~(\ref{po}) is the decay
into two fermions. Another source of damping, multiple spin excitation
scattering, is considered phenomenologically by adding the small
artificial broadening $-2\eta\omega_{\bf k}$, $\eta=0.02t$ to ${\rm
Im}\Pi({\bf k}\omega)$. The broadenings $-\eta$ is also added to ${\rm
Im}\Sigma({\bf k}\omega)$ to widen narrow lines and to stabilize the
iteration procedure.

The same derivation for the transversal Green's function gives
$\langle\!\langle s_{\bf k}^{-1}\big|s_{\bf k}^{+1}\rangle\!
\rangle=2D({\bf k}\omega)$ indicating that the used approach retains
properly the rotation symmetry of spin components in the paramagnetic
state.

For low $x$ and $T$ the bandwidth of the dispersion $\varepsilon_{\bf
k}$ is small in comparison with $8t$, the bandwidth of uncorrelated
electrons. This is a manifestation of the band narrowing in the
antiferromagnetic surrounding.

Equations~(\ref{se})--(\ref{po}) form a closed set which was solved by
iteration for the parameters $t=0.5$~eV, $J=0.1$~eV corresponding to
cuprates \cite{Mcmahan}.

\begin{figure}
\centerline{\includegraphics[width=7cm]{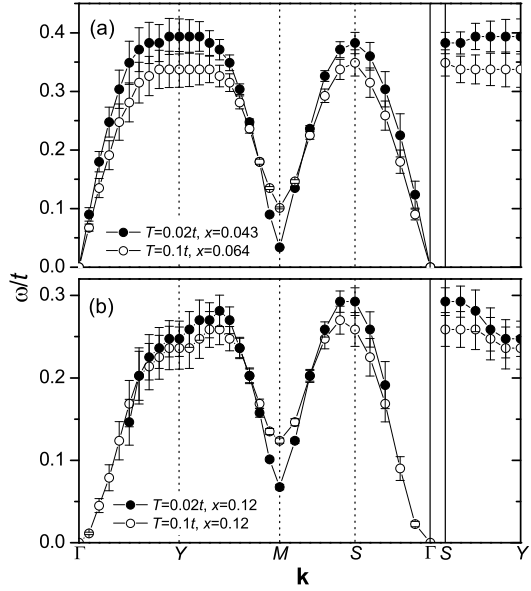}}
\caption{\label{Fig_i}The dispersion of spin excitations. Vertical bars
show decay widths $|{\rm Im}\Pi({\bf k}\omega)|/(2\omega_{\bf k})$.
Points $Y$, $M$ and $S$ correspond to ${\bf k}=(0,\pi)$, $(\pi,\pi)$
and $(\pi/2,\pi/2)$, respectively.}
\end{figure}

The frequencies of spin excitations satisfy the equation
\begin{equation}
\omega^2-{\rm Re}\Pi({\bf k}\omega)-\omega^2_{\bf k}=0 \label{sef}
\end{equation}
[see Eq.~(\ref{se})]. For low $x$ and $T$ their dispersion is close to
the dispersion of spin waves (see Fig.~\ref{Fig_i}a). The main
difference is the spin gap at $(\pi,\pi)$ the magnitude of which grows
with $x$ and $T$ (Fig.~\ref{Fig_i}b).
In an infinite crystal this gap is directly connected with the spin
correlation length $\xi$. Indeed, for large distances and low $T$ we
find
\begin{eqnarray}
\left\langle s^z_{\bf l}s^z_{\bf 0}\right\rangle&=&N^{-1}\sum_{\bf
k}{\rm e}^{i\bf kl}\int_0^\infty
d\omega\coth\left(\frac{\omega}{2T}\right) B({\bf k}\omega)\nonumber\\
&\propto&{\rm e}^{i\bf Ql}(\xi/|{\bf l}|)^{1/2}{\rm e}^{-|{\bf
l}|/\xi},\label{spincor}
\end{eqnarray}
where ${\bf Q}=(\pi,\pi)$ and $\xi=a/(2\sqrt{\Delta})$ with the
intersite distance $a$. For low $x$ we found that $\Delta\approx 0.2x$
and consequently $\xi\approx a/\sqrt{x}$. This relation has been
experimentally observed in La$_{2-x}$Sr$_x$CuO$_4$ \cite{Keimer}.

As seen from Fig.~\ref{Fig_i}b, with growing $x$ the spin excitation
branch is destroyed in some region around the $\Gamma$ point --
Eq.~(\ref{sef}) has no solution for real $\omega$ due to large negative
${\rm Re}\Pi({\bf k}\omega)$. For fixed $T$ the size of this region
grows with $x$. In the considered model with rise of $T$ the branch is
recovered in this region. This is a consequence of the temperature
broadening of the quasiparticle peak in the hole spectrum. With this
broadening $|{\rm Re}\Pi|$ becomes smaller and Eq.~(\ref{sef}) has
again a real solution.

\begin{figure}
\centerline{\includegraphics[width=7.5cm]{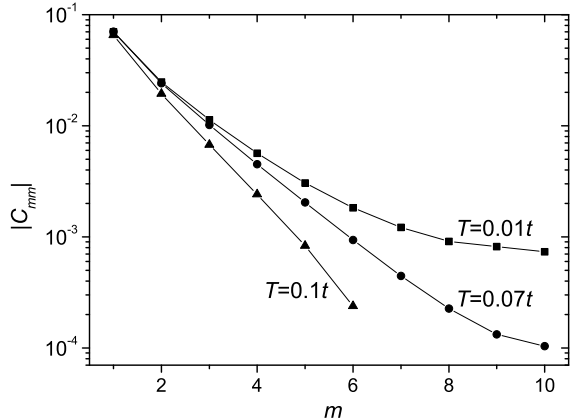}}
\caption{\label{Fig_ii}Spin correlations along the diagonal [i.e.,
${\bf l}=(m,m)$] of the crystal for $x=0.12$. The respective
temperatures are indicated near the curves.}
\end{figure}

An example of the variation of the spin correlations
$C_{mn}=\left\langle s^z_{\bf l}s^z_{\bf 0}\right \rangle$, ${\bf
l}=(m,n)$ with distance is given in Fig.~\ref{Fig_ii}.
For large enough $x$ and $T$ the correlations decay exponentially with
distance in the considered finite lattice. As mentioned, the method
used has no preset magnetic ordering. The character of the ordering is
determined in the course of the calculations. As seen from
Fig.~\ref{Fig_ii}, only the short-range antiferromagnetic ordering was
found in our calculations. Stripes or other types of the phase
separation were not revealed. Conceivably such phase separations are
not connected with the strong electron correlations described by the
$t$-$J$ model.

\begin{figure}
\centerline{\includegraphics[width=6.5cm]{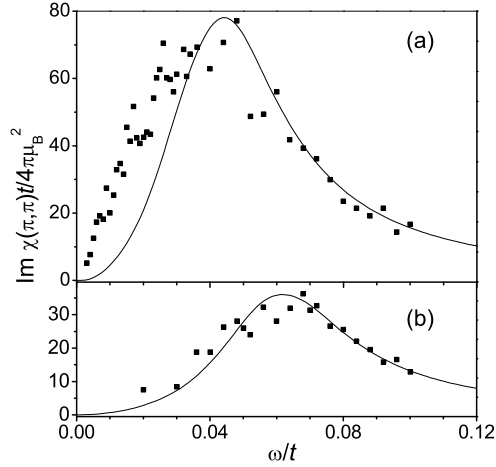}}
\caption{\label{Fig_iii}The
imaginary part of the spin susceptibility for ${\bf k}=(\pi,\pi)$.
Curves demonstrate calculated results for $T=0.02t\approx 116$~K,
$x=0.043$ (a) and 0.08 (b). Squares are experimental results obtained
in normal-state YBa$_2$Cu$_3$O$_{7-y}$ at $T=100$~K for $y=0.5$ (a) and
0.17 (b) \protect\cite{Fong}.}
\end{figure}

The magnetic susceptibility is connected with the spin Green's function
(\ref{se}) by the relation $\chi^z({\bf k}\omega)=-4\mu_B^2 D({\bf
k}\omega)$, where $\mu_B$ is the Bohr magneton. Experiments on
inelastic neutron scattering give information on the susceptibility
which can be directly compared with the calculated results. Such
comparison is carried out in Fig.~\ref{Fig_iii}.
YBa$_2$Cu$_3$O$_{7-y}$ is a bilayer crystal and the symmetry allows one
to divide the susceptibility into odd and even parts. For the
antiferromagnetic intrabilayer coupling the odd part can be compared
with the calculated results. The oxygen deficiencies $y=0.5$ and 0.17
in the experimental data in Fig.~\ref{Fig_iii} correspond to the hole
concentrations $x=0.05$ and 0.11, respectively \cite{Tallon}. As seen
from Fig.~\ref{Fig_iii}, the calculated data reproduce correctly the
frequency dependence of the susceptibility, the values of the frequency
for which ${\rm Im}\chi(\pi,\pi)$ reaches maximum and their evolution
with doping. The growth of the frequency of the maximum with $x$
reflects the respective increase of the spin gap. We notice also that
the calculated temperature variation of the susceptibility is in good
agreement with experiment. In absolute units the calculated maxima of
${\rm Im}\chi(\pi,\pi)$ are $1.5-2$ times larger than the experimental
values which is connected with some difference in decay widths of spin
excitations.

\begin{figure}
\centerline{\includegraphics[width=7cm]{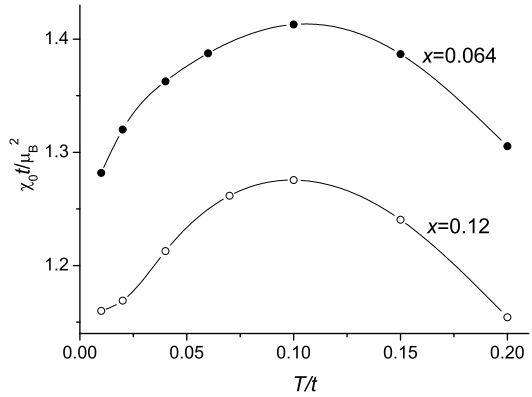}}
\caption{\label{Fig_iv}
The uniform static spin susceptibility vs.\ $T$.}
\end{figure}

The temperature dependence of the uniform static spin susceptibility
$\chi_0=\chi({\bf k}\rightarrow 0,\omega=0)$ is shown in
Fig.~\ref{Fig_iv}.
The calculated values lie in the range 2-2.6~eV$^{-1}$ which is close
to the values 1.9-2.6~eV$^{-1}$ obtained for YBa$_2$Cu$_3$O$_{7-y}$
\cite{Barzykin}. The dependence $\chi_0(T)$ has a maximum and its
temperature $T_m$ grows with decreasing $x$. Analogous behavior is
observed in cuprates for large enough $x$ \cite{Johnston,Takigawa}. In
Fig.~\ref{Fig_iv} $T_m\approx 600$~K which is close to $T_m$ observed
in La$_{2-x}$Sr$_x$CuO$_4$ for comparable $x$ \cite{Johnston}. As
known, in the undoped antiferromagnet $T_m\approx J$ \cite{Manousakis}.
On the high-temperature side $\chi_0(T)$ tends to the Curie-Weiss
dependence $1/T$. The decrease of $\chi_0$ below $T_m$ is sometimes
considered as the manifestation of the spin gap. In our opinion this
statement is incorrect. For moderate $x$ and $T$ the long-wavelength
part of the spin excitation spectrum does not feel the gap at
$(\pi,\pi)$. For small but finite values of {\bf k} $\chi({\bf
k},0)\propto \int_{-\infty}^\infty d\omega' B({\bf k}\omega')/\omega'$.
The function $B({\bf k}\omega')$ has a maximum which is shifted to
lower frequencies and loses its intensity with increasing $T$ for such
wave vectors. In the above integral the maximum is superimposed with
the decreasing function $1/\omega'$ which finally leads to the
nonmonotonic behavior of $\chi_0(T)$.

As seen from Fig.~\ref{Fig_iv}, the two curves for the different $x$
are very close in shape and can be superposed by scaling to the same
values of maximum $\chi_0$ and $T_m$. Analogous scaling was observed in
La$_{2-x}$Sr$_x$CuO$_4$ \cite{Johnston}. As follows from the above
discussion, the source of this scaling is that holes and temperature
fluctuations lead in a similar manner to the softening of the maximum
in $B({\bf k}\omega')$ for long wavelengths.

\begin{figure}
\centerline{\includegraphics[width=8.5cm]{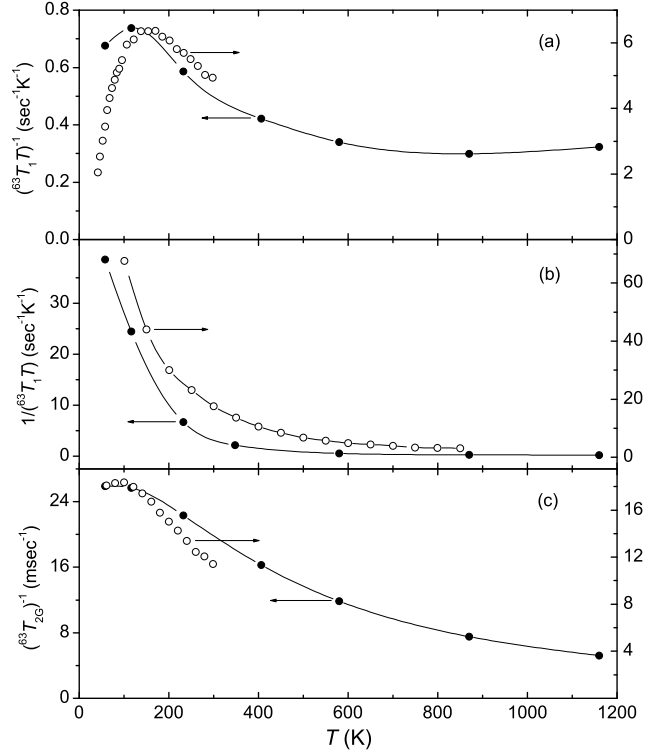}}
\caption{\label{Fig_v}The temperature dependencies of the spin-lattice
relaxation and spin-echo decay rates at Cu sites. Open circles with
right axes represent experimental results, filled circles with left
axes are our calculations. (a,c) calculations for ${\bf H\|c}$ and
$x=0.12$, measurements in YBa$_2$Cu$_3$O$_{6.63}
$\protect\cite{Takigawa} ($x\approx 0.1$ \protect\cite{Tallon}). (b)
calculations for nonoriented configuration with $x=0.043$, measurements
in La$_{1.96}$Sr$_{0.04}$CuO$_4$ \protect\cite{Imai}.}
\end{figure}

The spin-lattice relaxation and spin-echo decay rates were calculated
with the use of the equations \cite{Barzykin}
\begin{eqnarray}
\frac{1}{^\alpha T_{1\beta}T}&=&\frac{1}{2\mu^2_BN}\sum_{\bf k}
 \,^\alpha\!
 F_\beta({\bf k})\frac{{\rm Im}\,\chi({\bf k}\omega)}{\omega},
 \quad\omega\rightarrow 0,\nonumber\\
\frac{1}{^{63}T_{2G}^2}&=&\frac{0.69}{128\mu_B^4}\Biggl\{
 \frac{1}{N}\sum_{\bf k}\,^{63}\!F_e^2({\bf k})\left[{\rm Re}\,
 \chi({\bf k}0)\right]^2 \label{nmr}\\
&-&\left[\frac{1}{N}\sum_{\bf k}\,^{63}\!F_e({\bf k})
 {\rm Re}\,\chi({\bf k}0)\right]^2\Biggr\}, \nonumber
\end{eqnarray}
where the indices $\alpha$ and $\beta$ in the form factors
$^\alpha\!F_\beta({\bf k})$ \cite{Barzykin} indicate the nucleus type
and the direction of the applied static magnetic field {\bf H},
respectively. $\alpha=63$ corresponds to Cu. The form factor
$^{63}\!F_e$ is the filter for the Cu spin-echo decay time
$^{63}T_{2G}$. Our calculated results and the respective experimental
data are given in Fig.~\ref{Fig_v}.
The calculations reproduce satisfactorily main peculiarities of the
temperature dependencies of the spin-lattice and spin-echo decay rates.
The growth of $(^{63}T_1T)^{-1}$ with decreasing $x$ is connected with
the increase of the spectral intensity of spin excitations near
$(\pi,\pi)$ which make the main contribution to this rate. For the same
$x$ $(^{63}T_1T)^{-1}$ is one order of magnitude larger than
$(^{17}T_1T)^{-1}$, the spin-lattice relaxation rate at O sites (not
shown here). This is a consequence of ${\rm Im}\chi$ which is strongly
peaked near $(\pi,\pi)$ and the form factors which test different {\bf
k} regions \cite{Barzykin}. The calculated spin-lattice relaxation
rates are smaller than the experimental values due to the approximation
made in the calculation of $D({\bf k}\omega)$ which somewhat
underestimates ${\rm Im}\chi$ at low frequencies.

For moderate $x$ with increasing $T$ the low-frequency region of ${\rm
Im}\,\chi({\bf k\approx Q})$, ${\bf Q}=(\pi,\pi)$ first grows due to
the temperature broadening of the maximum in its frequency dependence
and then decreases due to the temperature growth of the spin gap. It is
the reason for the nonmonotonic behavior of $(^{63}T_1T)^{-1}$ in
Fig.~\ref{Fig_v}a. For moderate $x$ and low $T$ the magnitude of the
spin gap is determined by the hole concentration and does not depend on
$T$. This temperature range corresponds to the growth stage in
Fig.~\ref{Fig_v}a. The independence of the gap from $T$ means that in
this temperature range also $\xi$ does not depend on temperature which
is a distinctive feature of the quantum disordered regime
\cite{Chubukov,Barzykin}. For temperatures above the maximum of
$(^{63}T_1T)^{-1}$ we found that $^{63}T_1T/^{63}T_{2G}\approx$ const
(see Fig.~\ref{Fig_v}c) and $\xi^{-1}\propto\sqrt{\Delta}\propto T$
which points to the quantum critical $z=1$ regime. These results are in
agreement with the phenomenological treatment of experiment in
YBa$_2$Cu$_3$O$_{7-y}$ carried out in Ref.~\cite{Barzykin} and the
temperature of the maximum in Fig.~\ref{Fig_v}a is close to $T_*$ of
that work.

For small $x$ the spin gap grows with temperature starting from low $T$
and $(^{63}T_1T)^{-1}$ decreases monotonously, as seen in
Fig.~\ref{Fig_v}b. Due to the form-factor the region near $(\pi,\pi)$
does not contribute to $(^{17}T_1T)^{-1}$. There is a cardinal
difference between the behavior of ${\rm Im}\,\chi$ for ${\bf k\approx
Q}$ and away from $(\pi,\pi)$. Due to the spin gap in the former case
the frequency of the maximum in ${\rm Im}\,\chi(\omega)$ increases with
$T$, while in the latter case, as mentioned above, it decreases. This
frequency softening leads to the growth of the low-frequency ${\rm
Im}\,\chi$ and $(^{17}T_1T)^{-1}$ at low $T$ with their saturation for
higher temperatures. Analogous behavior is observed in experiment
\cite{Takigawa}.

This work was supported by the ESF grant No.~4022.

\end{document}